\documentclass[11pt]{article}
\usepackage[utf8]{inputenc} 
\usepackage[letterpaper,margin=1in]{geometry}
\usepackage{setspace} 
\usepackage{graphicx}   
\usepackage[format=plain,labelfont={bf,it},
            font=it]{caption}
\usepackage[section,above,below]{placeins} 
\usepackage{fancyvrb}  
\usepackage{textcomp}
\usepackage{gensymb} 
\usepackage{amssymb} 
\usepackage{bm}         
\usepackage{comment}
\usepackage{enumitem}
\usepackage{hyperref}  
\usepackage[toc,page]{appendix} 

\usepackage{csquotes}  
\MakeOuterQuote{"}  
\usepackage{array}
\usepackage{tabu}
\usepackage{ amssymb, amsmath}
\usepackage{listings}
\def\therefore{\boldsymbol{\text{ }
\leavevmode
\lower0.4ex\hbox{$\cdot$}
\kern-0.65em\raise0.7ex\hbox{$\cdot$}
\kern-0.65em\lower0.4ex\hbox{$\cdot$}
\thinspace\text{ }}}
\mathchardef\-="2D

\hypersetup{    
    colorlinks=true,
    linkcolor=blue,
    filecolor=magenta,      
    urlcolor=cyan,
}
\fvset{numbers=left,fontsize=\footnotesize,obeytabs,frame=single}  

\usepackage{caption}
\usepackage{subcaption}

\usepackage[style=numeric,
    sorting=nty,
    citestyle=numeric-comp,
    language=american,
    sortcase=false,
    sortcites=true,
    backend=biber]{biblatex}   
\usepackage{float}
\newcommand\figref[1]{Figure~\ref{fig:#1}}

\newcommand\eqnref[1]{Eq.~\ref{eqn:#1}}

\usepackage{placeins}

\usepackage{fancyvrb}   
\fvset{numbers=left,fontsize=\footnotesize,obeytabs,frame=single} 
\DeclareUnicodeCharacter{2212}{-} 

\title{Semiconductor-like Optical Properties Unveiled by Modeling of Short-Period Aluminum Oxide-Copper Multi-Layered Nanocomposites Deposited by Sputtering Atomic Layer Augmented Deposition (SALAD)}
\author{\textit{Søren A. Tornøe*, Jacob H. Sands**, and Nobuhiko P. Kobayashi*†}}

\begin{document}

\noindent \textbf{Semiconductor-like Optical Properties Unveiled by Modeling of Short-Period Aluminum Oxide-Copper Multi-Layered Nanocomposites Deposited by Sputtering Atomic Layer Augmented Deposition (SALAD)}
\newline \newline
\noindent \textit{Søren A. Tornøe*, Jacob H. Sands**, and Nobuhiko P. Kobayashi*†}
\newline \newline
\noindent *Søren A. Tornøe and Nobuhiko P. Kobayashi 
\noindent Nanostructured Energy Conversion Technology and Research (NECTAR), Department of Electrical and Computer Engineering, Baskin School of Engineering, University of California Santa Cruz, California 95064, United States of America
\newline \newline
\noindent **Jacob H. Sands
University of California Observatories (UCO), University of California Santa Cruz, 1156 High Street Santa Cruz, California 95064, United States of America
\newline \newline
†Corresponding author: E-mail: nkobayas@ucsc.edu

\section*{Keywords:}
\noindent sputtering, atomic layer deposition, copper, aluminum oxide, nanocomposites, thin film, transfer-matrix, index of refraction, dispersion integral

\section*{Abstract}

\noindent \par In our previous publication, we discussed the optical properties of multi-layered nanocomposite samples consisting of 300 pairs of an AlO\textsubscript{x} layer and a Cu layer produced from sputtering atomic layer augmented deposition (SALAD). These samples displayed unusual spectral reflectance that could not be well described via conventional means and could not be explained as an interpolation of the constituent materials. Computational models were then needed to both describe the results that were obtained and to predict the sample’s other material properties. This process was approached by using the transfer-matrix method to describe the spectral reflectance based on the constituent materials, and then the dispersion method was used to find the composite index of refraction for the samples. The models successfully described pure Cu but encountered minor difficulties when it came to AlO\textsubscript{x}, as such, more research is needed to either properly model non-holomorphic materials like AlO\textsubscript{x} with the current algorithm or to find a better model that can adequately describe both Cu and AlO\textsubscript{x}. The dispersion model implemented predicts that the samples, composed of many thin film layers of metal and dielectric, would display semiconductor-like behavior. Therefore, this indicates that SALAD’s combination of sputtering and ALD permits the generation of novel metal-dielectric based semiconductors.

\section{Introduction}

\noindent \par In our previous study, sputtering (SPU) and atomic layer deposition (ALD) were combined to establish a distinctive capability of depositing thin films; referred to as Sputtering Atomic Layer Augmented Deposition (SALAD) \cite{SALAD}. SALAD allows us to deposit complex multi-layered structures while offering the benefits of both sputtering and ALD within the same vacuum chamber. SALAD was used to prepare samples – short-period AlO\textsubscript{x}-Cu multi-layered nanocomposites – comprising 300 pairs with each pair made of a layer of ALD deposited aluminum oxide (AlO\textsubscript{x}) with a nominal thickness of $39\,nm$ and a layer of sputtering deposited copper (Cu) with nominal thickness ranging between $15$ to $84\,nm$ \cite{SALAD}. When tested through spectroscopic ellipsometry, the samples displayed spectral reflectance that did not appear to be inherent to a simple interpolation of the constituent materials (i.e., AlO\textsubscript{x} and Cu). This statement held particularly true in the spectral range of low ultraviolet light to near the end of the visible spectrum (i.e., the range of wavelengths from $200\,nm$ to $600\,nm$). The implication, it would seem, is that these samples have unusual optical properties to which conventional views such as the effective media approximation simply do not apply \cite{SALAD}. This peculiar observation, therefore, calls for further investigation for better understanding of these distinctive optical properties.

\noindent \par The goal of the study presented in this paper is to explore two modeling methods to illustrate the spectral reflectance experimentally obtained from a series of short-period AlO\textsubscript{x}-Cu multi-layered nanocomposites prepared by SALAD and to obtain a composite index of refraction for the samples from said spectral reflectance, respectively. The first portion of the study focuses on calculating the spectral reflectance of the samples via the transfer-matrix method which approximates all forward and backward light wave propagation in the samples into one 4x4 matrix \cite{OIRD}, provided that the index of refraction and thicknesses of the constituent materials (i.e., AlO\textsubscript{x} and Cu) are known. The second portion of the study uses the dispersion method to obtain the composite index of refraction from experimental spectral reflectance \cite{LiF_Opt}. Together, these two complementary methods hold promise in revealing electronic properties, via optical measurements, of samples for which direct electrical characterization through traditional means is a non-trivial task. We feel confident in our model’s prediction then, that alternating thin films of a metal and a dielectric material used in the SALAD samples would exhibit optical properties qualitatively comparable to those of semiconductors. 

\section{Experiment}

\noindent \par A series of short-period AlO\textsubscript{x}-Cu multi-layered nanocomposites were prepared by SALAD. While the details of the samples were reported previously \cite{SALAD}, a summary is provided as follows. AlO\textsubscript{x} was chosen due to its maturity as a dielectric material routinely deposited by ALD (ALD-AlO\textsubscript{x}), and Cu was chosen as a metallic material frequently deposited by SPU (SPU-Cu). Additionally, the oxides (e.g., cuprous and cupric oxides) that could be formed from Cu are easily distinguishable from AlO\textsubscript{x}, making the theoretical analysis simpler.

\noindent \par The deposition rate of ALD-AlO\textsubscript{x} ($\Gamma_{ALD}$) was self-limiting to $0.13\,nm/cycle$. The deposition rate of SPU-Cu ($\Gamma_{SPU}$) was tuned to be $0.04\,[nm/s]$ allowing for the nominal content of Cu sub-layer to be controlled by specifying a unique time duration $t_{SPU}\,s$ for SPU-Cu. A single SALAD cycle consisted of an $Al(CH_3)_3$ pulse, a purge for the remaining $Al(CH_3)_3$, a $H_2O$ pulse, a purge for the remaining $H_2O$, followed by a sputtering of Cu for a desired $t_{SPU}$. A sample was produced by repeating a single SALAD cycle 300 times for a given $t_{SPU}$, and consequently, a series of samples were prepared with varied Cu content represented by $\tau$. All depositions were carried out at $150\degree C$ at pressure in the range of $0.14-0.24\,torr$ using argon for both ALD and SPU (i.e., argon carrier/purge gas for ALD and argon gas plasma in SPU).

\noindent \par Spectroscopic reflectometry was performed on the samples using the FilmTek 4000 spectroscopic reflectometry/ellipsometry equipment where the change in s-polarized light was measured to determine reflectance with a given angle of incident around $70\degree$. Collecting data through spectroscopic reflectometry was chosen over directly obtaining the composite index of refraction, $n$ and $k$ values, through spectroscopic ellipsometry as the ellipsometry modeling required structural details about the samples that were simply not adequately obtainable. Spectroscopic reflectometry was carried out in the spectral range of $0.48-6.0\,eV$; however, our focus is paid mostly to $1.6-6.0\,eV$ as, in this spectral range, the samples were clearly differentiated from each other. 

\noindent \par The spectral reflectance of the samples are presented in \figref{SALAD_graphs}. All the samples commonly depict a broad peak – convex dome-like shape – located in the spectral range from $2.2\,eV$ to $5.6\,eV$. The onset of each convex dome-like shape appears to be red-shifted as $\tau$ increases, and the high-energy side of each convex dome-like shape seems to contain 2-3 satellite peaks. Similarly, the samples generally display an increase in reflectance below $2.2\,eV$, showing a valley in reflectance at around $2\,eV$, which is especially notable as this feature does not appear to be a property native to Cu nor AlO\textsubscript{x}. In general, the reflectance of Cu nearly reaches its maximum at $\sim 2\,eV$ \cite{QuerryCu} and that of AlO\textsubscript{x} gradually decreases as energy decreases in the spectral range around $2\,{eV}$ \cite{QuerryCu}. In other words, AlO\textsubscript{x} and Cu are expected to show predictable reflectance spectra if separately analyzed; thus, the peculiar features seen in \figref{SALAD_graphs} are due in part to the unique way, offered by SALAD, by which AlO\textsubscript{x} and Cu are combined. For instance, the number of interfaces (i.e., 300) in the samples is much larger than those normally found in conventional optical coatings. In addition, nominal thicknesses of the sub-layers (i.e., AlO\textsubscript{x} and Cu) that form a single repeating unit are much thinner than those generally used in traditional optical coatings. 

\noindent \par Absorption (i.e., a decrease in reflectance) associated with bulk plasmon polariton of Cu would arise at $\sim 2.3\,eV$ and continue to $\sim 4\,eV$ \cite{QuerryCu} if Cu dominantly played a role in constructing the spectra in \figref{SALAD_graphs}; however, all the spectra commonly show the convex dome-like shape with a sharp increase, in particular for $\tau$ in the range of $0.48 \sim 0.68$, in reflectance that starts in the range of $1.8 \sim 2.4\,eV$. The aforementioned increase gives rise to a pronounced dip in reflectance in the same energy range, resulting in a drop below what is expected for AlO\textsubscript{x}. For each $\tau$, reflectance continues to drop as energy increases once the first peak is reached (the first peak, for instance, $\tau\,=\,0.68$ is indicated by the black arrow). Beyond the first peak, characteristic satellite peaks, for instance satellite peaks denoted by the three red arrows seen in the range of $3.0\,eV$ to $5.60\,eV$ for $\tau\,=\,0.68$ are present, at least one of which could be associated with surface plasmon polariton (SPP) generated at the interfaces between ALD-AlO\textsubscript{x} and SPU-Cu. These satellite peaks could also mirror the complex behavior including the bulk plasmon resonances of Cu above the plasma frequency at around $2\,eV$. Apart from plasmon polaritons, there may be a form of varying codominance between AlO\textsubscript{x} and Cu occurring across the range of energy where SPU-Cu becomes dominant in the higher energy range when $\tau$ is above $0.38$. In addition, the samples may be regarded as metal-dielectric superlattices such as hyperbolic metamaterials that give rise to enormously high photonic densities of states over a broad spectral range \cite{HyperbolicMeta}. The term “short-period” in the name of the samples – the short-period AlO\textsubscript{x}-Cu multi-layered nanocomposites – is in reference to the expression conventionally used in describing superlattices of semiconductor layers with thicknesses in the range of few nanometers – short-period superlattices – where the bound-state energy-level splits because of the coupling of electronic states of well layers through barrier layers \cite{SLatt1}\cite{SLatt2}\cite{SLatt3}, which is expected to produce optical properties vastly different from the constituent materials.

\section{Modeling}

\noindent \par With all these possible explanations to the peculiar spectral reflectance presented in \figref{SALAD_graphs}, we proceeded to analyze the spectral reflectance using the following two complementary methods: the transfer-matrix method and the dispersion method to extract further insights from the spectral reflectance. All of the modeling described in the following section was carried out using GNU Octave \cite{octave}. 

\subsection{Obtaining Spectral Reflectance Using The Transfer-Matrix Method}

\noindent \par To understand the spectral reflectance experimentally obtained and shown in \figref{SALAD_graphs}, spectral reflectance of the short-period AlO\textsubscript{x}-Cu multi-layered nanocomposites samples were modeled using the transfer-matrix method. The transfer matrix method relies on the complex index of refraction for each of the sub-layer materials at a given vacuum wavelength, $\lambda_0$. For AlO\textsubscript{x} sub-layers and the SiO\textsubscript{2} substrate, the Sellmeier equation was implemented with coefficients found in literatures \cite{CRC}\cite{MatrlSci}. 

\noindent \par While the Sellmeier equation is viable for transparent dielectric materials such as AlO\textsubscript{x}, there exist no empirical means to use the equation to model  index of refraction of metals such as Cu; thus, a mathematical expression was derived by spline fitting existing numerical data sets to create a mathematical expression of energy dependent index of refraction for Cu \cite{QuerryCu}. The resulting mathematical expression works satisfactorily so long as the chosen energy is within the spectral range of interest of $0.42\,eV \sim 6.0\,eV$.

\noindent \par The structure shown in \figref{layers} that factors in the presence of all of the 300 interfaces and specific sub-layer thicknesses with the variations in $\tau$ was used in the modeling, revealing how complex the calculations become and that calculating the reflectance of a multi-layered structure consisting of 300 layers verges on overwhelmingly onerous even though the transfer-matrix method allows us to simplify the calculations for any layered material while taking into consideration forward and backward propagating light waves.

\noindent \par Two assumptions were made with regards to the modeling of spectral reflectance by using the transfer-matrix method. The first assumption was that the morphological roughness present at the 300 interfaces of the samples is negligible; thus, the interfaces are treated as geometric boundaries at which one material (e.g., AlO\textsubscript{x}) changes to the other (e.g., Cu) abruptly. In other words, there is no compositional mixing between sub-layers (i.e., $n$ and $k$ change abruptly from those of AlO\textsubscript{x} to those of Cu and so on). The second assumption simply states that all layers no matter how thin are complete and homogeneous with their intended nominal thicknesses, being consistent with the first assumption and ensuring that the index of refraction within each sub-layer is isotropic \cite{SALAD}.

\noindent \par Focus was paid primarily to the s-polarized version of the equations, since the spectral reflectance presented in \figref{SALAD_graphs} were experimentally collected using s-polarized light. The transmission and the reflection coefficients, $t_s^{(m,m+1)}$ and $r_s^{(m,m+1)}$ respectively, of a homogeneous $m+1^{th}$ sub-layer with respect to those of a $m^{th}$ sub-layer are calculated using

\begin{equation}\label{eqn:ts}
    t_s^{(m,m+1)} = \frac{2\,n_m\, cos(\theta_m)}{n_{m}\, cos(\theta_m)+n_{m+1}\, cos(\theta_{m+1})}
\end{equation}

\begin{equation}\label{eqn:rs}
    r_s^{(m,m+1)} = \frac{n_m\, cos(\theta_m)-n_{m+1}\, cos(\theta_{m+1})}{n_m\, cos(\theta_m)+n_{m+1}\, cos(\theta_{m+1}})
\end{equation}

\noindent assuming that the incident light was s-polarized \cite{OIRD}. In \eqnref{ts} and \eqnref{rs}, $n_m$ represents the index of refraction of an $m^{th}$ sub-layer from which light exits, $\theta_m$ is the incident angle of the light at the interface between the two sub-layers leaving the $m^{th}$ sub-layer, $n_{m+1}$ is the index of refraction of the $m+1^{th}$ sub-layer that the light enters, and $\theta_{m+1}$ – calculated through Snell’s law – is the incident angle of the light as it enters $m+1^{th}$ sub-layer. \eqnref{ts} and \eqnref{rs} only account for light propagating in the forward direction (i.e., light with directions having positive components when its trajectory is projected on the z-direction in \figref{layers}); however, in a multi-layered system such as the one in \figref{layers}, there is back-propagating light as seen in \figref{layers}, (i.e., each interface virtually produces back-propagating light that then individually interacts with each layer it comes into contact with), posing an issue of infinite calculations.

\noindent \par The remedy to the aforementioned issue inherent to multi-layered systems is the implementation of the transmission or interface matrix

\begin{equation}\label{eqn:matD}
    D_{p|s}^{(m,m+1)} = \frac{1}{t_{p|s}^{(m,m+1)}} 
    \begin{bmatrix}
    1 & r_{p|s}^{(m,m+1)} \\
    r_{p|s}^{(m,m+1)} & 1
    \end{bmatrix}
\end{equation}

\noindent represented by $D_{p|s}^{(m,m+1)}$ \cite{OIRD}. It should be noted that the general form of \eqnref{matD} does not change if the light is p-polarized or s-polarized, though the transmissivity, $t_{p|s}^{(m,m+1)}$, and the reflection coefficient, $r_{p|s}^{(m,m+1)}$, do have p-polarized and s-polarized specific forms. A $D_{p|s}^{(m,m+1)}$ matrix is generated for each layer (i.e., in our case, 300 matrices need to be generated), and multiplying them together effectively estimates the total light transmitted through all 300 layers in the samples and the total light reflected at the interface plus all of the light associated with reflections occurring at all the interfaces.

\noindent \par Another component that needed to be factored in is that light attenuates while inside a material losing its intensity relative to the distance, $d_m$ that the light has to travel (i.e., absorption of light). Thus, affecting the distance to which light can penetrate the sample and the intensity of the total reflected light. The incident light suffers attenuation relative to $e^{i\gamma_m}$, where $\gamma_m$ represents the propagation of the plane-wave in the z-direction as depicted in \figref{layers}. Readjusting to put the attenuation in terms of all the light passing through the material in both directions creates

\begin{equation}\label{eqn:matP}
    P_{p|s}^{(m)} = \begin{bmatrix}
    e^{-i\, \gamma_m} & 0 \\
    0 & e^{i\, \gamma_m}
    \end{bmatrix}
\end{equation}

\begin{equation}\label{eqn:gamma}
    \gamma_m = 2\, \pi\, n_m\, cos(\theta_m)\frac{d_m}{\lambda_0}
\end{equation}

\noindent referred to as the propagation or layer matrix \cite{OIRD}. The variable $\theta_m$ in \eqnref{gamma} is the incident
angle of the light at that sub-layer’s surface in radians. The transfer-matrix, $M$, then is the multiplication of the transmission matrices and propagating matrices in the order of the sub-layers illustrated in the structure in Fig. 2.

\begin{equation}\label{eqn:matM}
    M = D_{p|s}^{(0,1)} \,
    P_{p|s}^{(1)}\, D_{p|s}^{(1,2)}\, 
    P_{p|s}^{(2)}\, (D_{p|s}^{(2,1)}\,
    P_{p|s}^{(1)}\, D_{p|s}^{(1,2)}\, 
    P_{p|s}^{(2)})^{299}\, 
    D_{p|s}^{(2,s)}
\end{equation}

\noindent \par \eqnref{matM} is the transfer-matrix generated for the sample consisting of 600 sub-layers. Wherein an m of 0, 1, 2, and s represents air, Cu sub-layers, AlO\textsubscript{x} sub-layers, and the $SiO_2$ substrate, respectively. \eqnref{matM} is further simplified by recognizing the presence of the repeating sub-layer pattern (i.e., the portion in the large parentheses in \eqnref{matM}) for every layer, reducing most of the equation to the aforementioned pattern raised to the 299th power. Relating the transfer matrix M with an input light ($E_{in}$), a reflected light ($E_r$), and a transmitted light ($E_t$); we get

\begin{equation}\label{eqn:totalWave}
    \begin{bmatrix}
    E_{in} \\ E_r
    \end{bmatrix} 
    = M \begin{bmatrix}
    E_t \\ 0
    \end{bmatrix}.
\end{equation}

\noindent Simplifying \eqnref{totalWave} yields the following relations,

\begin{equation}
    E_{in} = M_{11} \,E_t
\end{equation}

\begin{equation}
    E_r = M_{21} \,E_t
\end{equation}

\begin{equation}\label{eqn:reflect}
    \frac{E_r}{E_{in}} = r = \frac{M_{21}}{M_{11}}
\end{equation}

\noindent \eqnref{reflect} thus indicates that the reflection coefficient r can be summarized as the $M21$ element of the transfer-matrix divided by the $M11$ element for a given wavelength $\lambda_0$ \cite{OIRD}. The absolute value of $r$ was squared to obtain the spectral reflectance, $R$. This calculation was then repeated every $1\,nm$ over the range of $190\,nm$ to $3\,\mu m$ to generate spectral reflectance for the samples with varied $\tau$.

\subsection{Recovering \textit{n} and \textit{k} Using the Dispersion Method}

\noindent \par The dispersion method, in short, is a computation that predicts the index of refraction of a material solely based on the material’s spectral reflectance and is even capable of predicting a composite index of refraction like those associated with our SALAD samples. The model implemented takes coefficients of reflection, places them in the polar coordinate plane, and uses a dispersion integral to determine the phase angle $\phi$ which in conjunction with the coefficients of reflection is then used to solve for $n$ and $k$ \cite{LiF_Opt}\cite{DispRelation}. However, there are challenges associated with this method as the dispersion integral requires, in principle, knowledge of the reflection coefficients from the entire energy domain (i.e., energy ranging from $0$ to $\infty\,eV$), although spectral reflectances obtained experimentally are always limited to a small portion of the domain. To properly implement the dispersion method, it is necessary to look into the details of the method’s derivation from reflectance to refractive index.

\noindent \par The complex reflection coefficient given normal incident light can be defined as

\begin{equation}\label{eqn:expRef}
    \sqrt{R} = \frac{n-i\,k-1}{n-i\,k+1} = |r|e^{i\,\phi}
\end{equation}

\noindent where $\phi\,[radians]$ is the phase angle associated with the coefficient of reflectance in the polar coordinate system \cite{LiF_Opt}. Solving \eqnref{expRef} for $n$ and $k$ individually results in

\begin{equation}\label{eqn:n}
    n = \frac{1-R}{1+R-2\,\sqrt{R}\,cos(\phi)}
\end{equation}

\begin{equation}\label{eqn:k}
    k = \frac{-2\,\sqrt{R}\,sin(\phi)}{1+R-2\,\sqrt{R}\,cos(\phi)}
\end{equation}

\noindent \cite{LiF_Opt}. In order to obtain n and k using \eqnref{n} and \eqnref{k} in reference to the spectral reflectance shown in \figref{SALAD_graphs}, $\phi$ needs to be calculated first, which is achieved by taking advantage of the dispersion theory as follows. First, by taking the natural log of \eqnref{expRef} to get

\begin{equation}
    \ln{\sqrt{R}} = \ln{r}+i\,\phi
\end{equation}

\noindent which places \eqnref{expRef} into the necessary format that allows the use of the dispersion theorem that states that the imaginary part, $\phi$, can be expressed as 

\begin{equation}
    \phi(\omega)=\frac{2\omega}{\pi} \int_{0}^{\infty} \frac{ln\,r(\omega_1)}{\omega^2-\omega_1^2} \,d\omega_1,
\end{equation}

\noindent the dispersion integral of the real part of the coefficient of reflection which depends on frequency $\omega$, over $\omega_1$ from $0$ to $\infty$ \cite{DispRelation}. The integral across the entire frequency spectrum poses an issue as it requires information outside the spectral range of the data used in the experiment – the boundary issue. The solution to the boundary issue was to separate the integral into four pieces

\begin{equation}\label{eqn:lower}
    \phi_1(\omega)=\frac{2\omega}{\pi} \int_{0}^{a} \frac{r_l}{\omega^2-\omega_1^2} \,d\omega_1
\end{equation}

\begin{equation}\label{eqn:a}
    \phi_2(\omega)=\lim_{b\to\omega}\frac{2\omega}{\pi} \int_{a}^{\omega-0.01} \frac{ln\,r(\omega_1)}{\omega^2-\omega_1^2} \,d\omega_1
\end{equation}

\begin{equation}\label{eqn:b}
     \phi_3(\omega)=\lim_{b\to\omega}\frac{2\omega}{\pi} \int_{\omega+0.01}^{b} \frac{ln\,r(\omega_1)}{\omega^2-\omega_1^2} \,d\omega_1
\end{equation}

\begin{equation}\label{eqn:upper}
    \phi_4(\omega)=\frac{2\omega}{\pi} \int_{b}^{\infty} \frac{r_u}{\omega^2-\omega_1^2} \,d\omega_1
\end{equation}

\noindent that allows the spectral range outside of what was collected in the experiment to be dealt with independently. The separation of the integral results in $\phi_1$ and $\phi_4$, each requiring a constant of integration – lower boundary $r_l$ and upper boundary $r_u$  respectively – that represent all possible $ln[r(\omega_1)]$ for that boundary region. Based on known data sets for Cu and AlOx, averages of the reflectance for the lower and upper out of bound regions were used to make initial estimates for the integration constants, $r_l$ and $r_u$, for the two constituent materials, Cu and AlO\textsubscript{x}, which yields $r_{lCu}$, $r_{uCu}$, $r_{lAlOx}$, and $r_{uAlOx}$. 

\noindent \par With these estimates as starting points, a calibration algorithm for both Cu and AlO\textsubscript{x} was developed to obtain $r_{lCu}$, $r_{uCu}$, $r_{lAlOx}$, and $r_{uAlOx}$. For Cu, the calibration algorithm tested all possible combinations of incrementing and decrementing $r_{lCu}$ and $r_{uCu}$ (i.e., $r_l$ and $r_u$ for Cu) in \eqnref{lower} and \eqnref{upper}. The results of the dispersion equations were added together and then n and k were calculated using \eqnref{n} and \eqnref{k} over the spectral range ($1.6\,eV$ to $6.0\,eV$) used in the experiment. The eight generated index spectra were compared to known n and k spectra for Cu. Subsequently, the algorithm identified the combination of $r_{lCu}$ and $r_{uCu}$ that was most similar to the known data set, and the process was repeated until the process converged on a set of $r_{lCu}$ and $r_{uCu}$ with sufficient accuracy in the desired energy range. The same procedure was used to calibrate $r_{lAlOx}$ and $r_{uAlOx}$ for AlO\textsubscript{x}. 

\noindent \par In the dispersion method, the samples were treated as a single layer made of a homogeneous material, the outcomes of the two calibrations separately done for Cu and AlO\textsubscript{x} needed to be fused back together by having the determined integration constants weighted and added based on the sample identifier $\tau$ that represents the atomic percent ratio of the two constituent materials (i.e., Cu and AlO\textsubscript{x}) as follows; 

\begin{equation}\label{lowr}
    r_{l} = r_{l_{Cu}}\,p + r_{l_{AlOx}}\,(1-p)
\end{equation}

\begin{equation}\label{uper}
    r_{u} = r_{u_{Cu}}\,p + r_{u_{AlOx}}\,(1-p),
\end{equation}

\noindent which effectively approximates the integration constants (i.e., $r_l$ and $r_u$) that represent the samples with varied $\tau$.

\noindent \par An additional separation of the integral (i.e., \eqnref{a} and \eqnref{b}) not associated with the boundary was implemented to prevent the integral from becoming undefined when $\omega$ would equal to $\omega_1$. Effectively, \eqnref{a} integrates along the range of the lower boundary $a$ – $200\,nm$ ($\sim 6.2\,eV$) – up to $\omega-0.01$, then \eqnref{b}. integrates from above $\omega+0.01$ to the upper bound at $2500\,nm$ ($\sim 0.5\,eV$); thus, resulting in the addition of $\phi_2$ and $\phi_3$. With all four integrals solved, $\phi_1$ through $\phi_4$ were added together and plugged into \eqnref{n} and \eqnref{k}. This process was repeated for all energies that fell within the spectral range of $1.6\,eV$ to $6.0\,eV$ used in the experiment.

\section{Results and Discussion}

\subsection{Results of the Transfer-Matrix Method}

\noindent \par
\begin{figure} [h]
    \centering
    \includegraphics[page=1, trim=3.8cm 0cm 3.8cm 0cm, clip, width=1\textwidth]{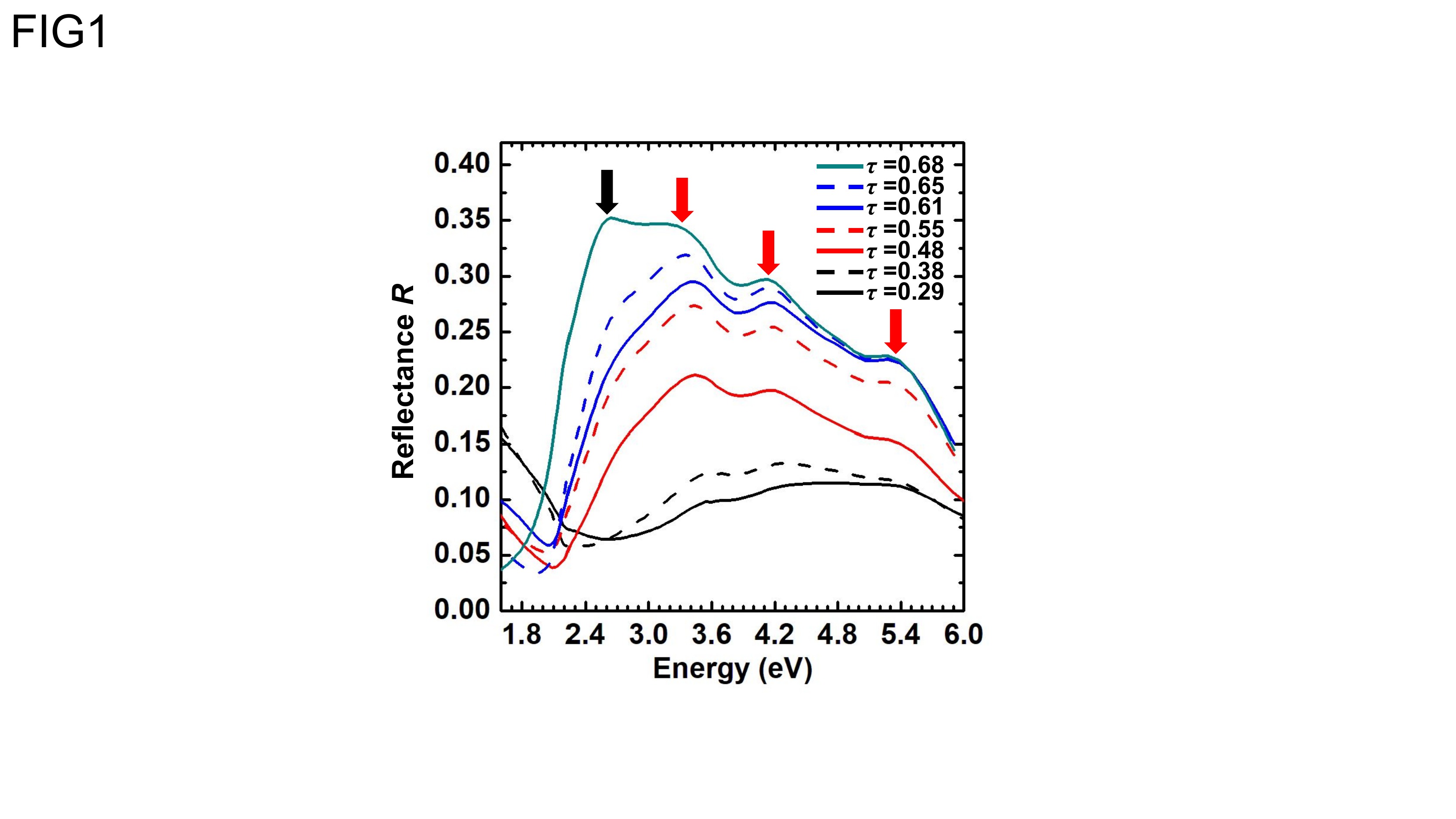}
    \captionsetup{labelformat=empty}\caption{}
    \label{fig:SALAD_graphs}
\end{figure}

\begin{figure} [h]
    \centering      
    \includegraphics[page=2, trim=3.8cm 0cm 3.8cm 0cm, clip, width=1\textwidth]{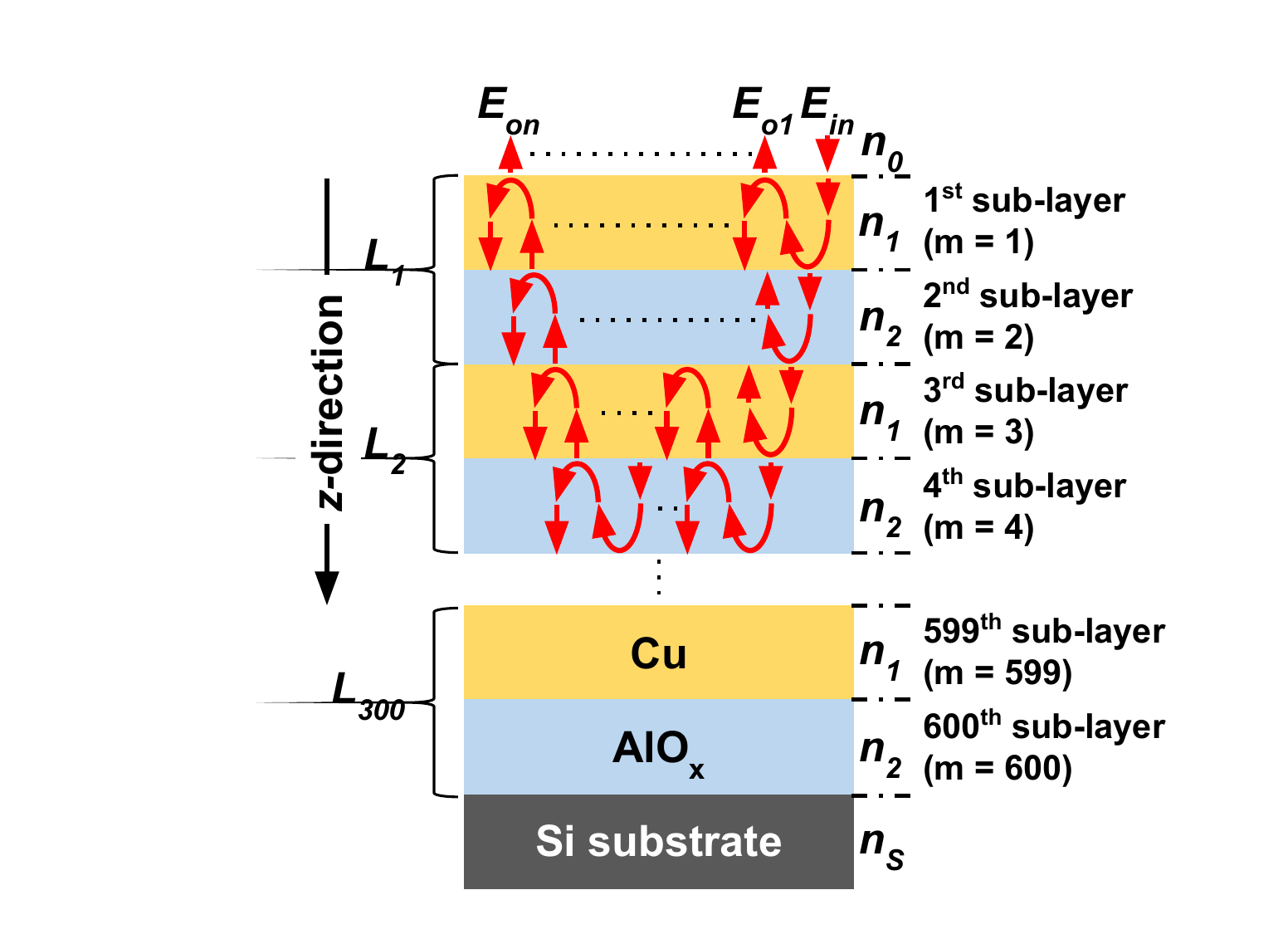}
    \captionsetup{labelformat=empty}\caption{}
    \label{fig:layers}
\end{figure}

\begin{figure} [h]
    \centering      
    \includegraphics[page=3, trim=3.8cm 0cm 3.8cm 0cm, clip, width=1\textwidth]{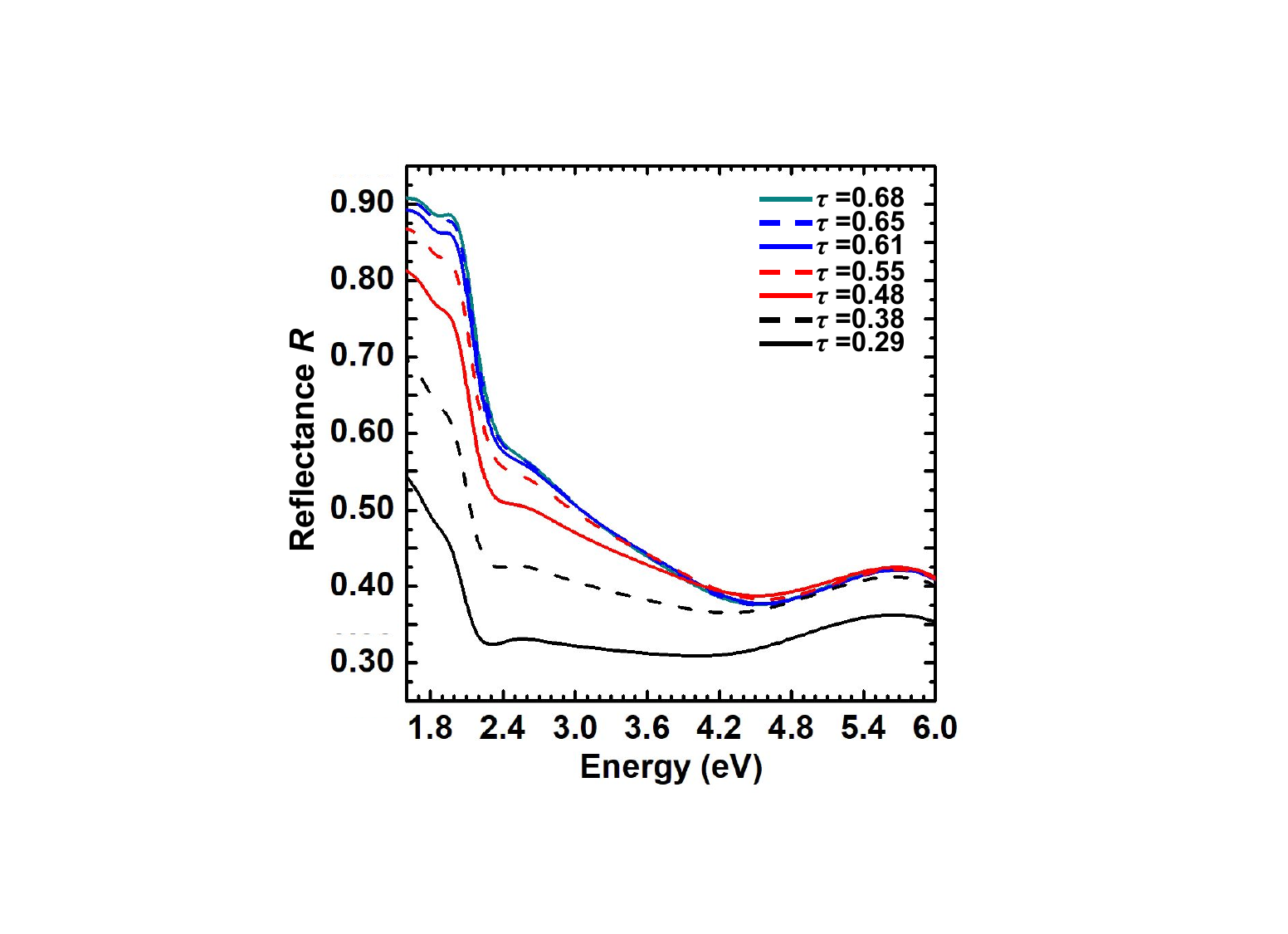}
    \captionsetup{labelformat=empty}\caption{}
    \label{fig:TrnMat_Results}
\end{figure}

 \figref{TrnMat_Results} shows spectral reflectance obtained by the transfer-matrix method. Comparing \figref{TrnMat_Results} with the spectral reflectance experimentally collected from the samples, \figref{SALAD_graphs}, it is immediately clear that the transfer-matrix method fails to yield the spectral reflectance of the samples. The only potential exception seems to be the sample with $\tau\,=\,0.29$ – the lowest Cu content.  The calculated spectrum for $\tau\,=\,0.29$ appears to show features qualitatively comparable to those seen in the spectrum of its counterpart in \figref{SALAD_graphs}, at least in terms of an overall shape. However, the convex dome is shifted primarily to the higher energy side in the calculated spectrum. Apart from the spectrum of $\tau\,=\,0.29$, there are clearly two major discrepancies between the spectra in \figref{SALAD_graphs} and those in \figref{TrnMat_Results}, which presumably are associated with the assumptions made in carrying out the transfer-matrix modeling. 

\noindent \par Most prominently, the satellite peaks seen as a common feature for the spectra with $\tau$ above $0.29$ in \figref{SALAD_graphs} do not appear in \figref{TrnMat_Results}, clearly suggesting that several proposed mechanisms mentioned earlier for explaining the presence of the satellite peaks, including absorption associated with surface plasmon polaritons, are indeed rational; in other words, these mechanisms are not included in the modeling. Therefore, the lack of satellite peaks appearing in \figref{TrnMat_Results}, is expected. Furthermore, all the spectra in \figref{TrnMat_Results} need to be shifted down by at least $25\%$ to be on the same scale as \figref{SALAD_graphs}, suggesting the contribution from Cu is overstated in the modeling, in other words, Cu is predicted to have a higher influence on the spectral reflectance than was observed, in particular, in the spectral range below the Cu plasma frequency at $\sim 2\,eV$ where it is amplified as in \figref{TrnMat_Results}. 

\noindent \par One of the main assumptions was that morphological roughness at the interfaces was negligible and that the transition in chemical composition at the interfaces occurred abruptly (i.e., there is no mixing of chemical elements at the interfaces). The samples were prepared by SALAD in which DC magnetron sputtering was used for the Cu layers and ALD was used for the AlO\textsubscript{x} layers. Sputtering frequently yields layers with substantially rough surfaces \cite{FundNano}, which implies that the implemented transfer-matrix model fails to accommodate the interface roughness when given a non-normal incident angle, drastically affecting the results from \eqnref{ts}, \eqnref{rs}, and \eqnref{gamma} further throwing off the results in \figref{TrnMat_Results}. In addition, some degree of diffusion is most likely to occur across the interfaces at elevated temperatures (Note: the samples were prepared at $150\degree C$). It is known that Cu is stabilized via the formation of copper aluminate spinel ($CuAl_2O_4$) and cuprous aluminate delafossite ($CuAlO_2$) when treated at high temperatures with the presence of $\gamma\- Al_2O_3$ \cite{HazardMat}, indicating that Cu is likely to have diffused and reacted with AlO\textsubscript{x} in the samples. Therefore, to replicate the spectral reflectance in \figref{SALAD_graphs} using the transfer-matrix method, it would be necessary then to account for a gradient of refractive indices across the interfaces of the two constituent materials. The second major assumption made was that even the thinnest SPU-Cu layers were continuous and free of holes. The nominal thickness of a single SPU-Cu layer in the samples is in the range of $0.05\,nm \sim 0.28\,nm$. Although direct assessment of ways by which these layers are incorporated in the samples using an analytical tool such as transmission electron microscope is expected to be extremely challenging if it is not impossible, it can be said, with a high degree of certainty, that some of these Cu layers, in particular, those with smaller $\tau$ are not continuous. Along similar lines, the assumption was also made that the transfer-matrix method is valid even when a single layer is much thinner than a quarter of the wavelengths examined in the modeling. Even though the transfer-matrix method has been successfully implemented in obtaining spectral reflectance of various optical thin film structures \cite{LiF_Opt}, it is quite possible that the method fails under these assumptions and is not valid for the short-period AlO\textsubscript{x}-Cu multi-layered nanocomposite samples.

\noindent \par It is already clear from the discussion above that the transfer-matrix method as it is implemented for the short-period AlO\textsubscript{x}-Cu multi-layered nanocomposite samples is currently incomplete. That said, with spectral reflectance of the samples being available, the dispersion method, as described in the following section, should be able to extract more insights from the samples by obtaining $n$ and $k$ spectra from the available spectral reflectance.

\subsection{Recovered \textit{n} and \textit{k} from Dispersion Method}

\noindent \par
\begin{figure} [h]
    \centering     
    \includegraphics[page=4, trim=3.8cm 0cm 3.8cm 0cm, clip, width=1\textwidth]{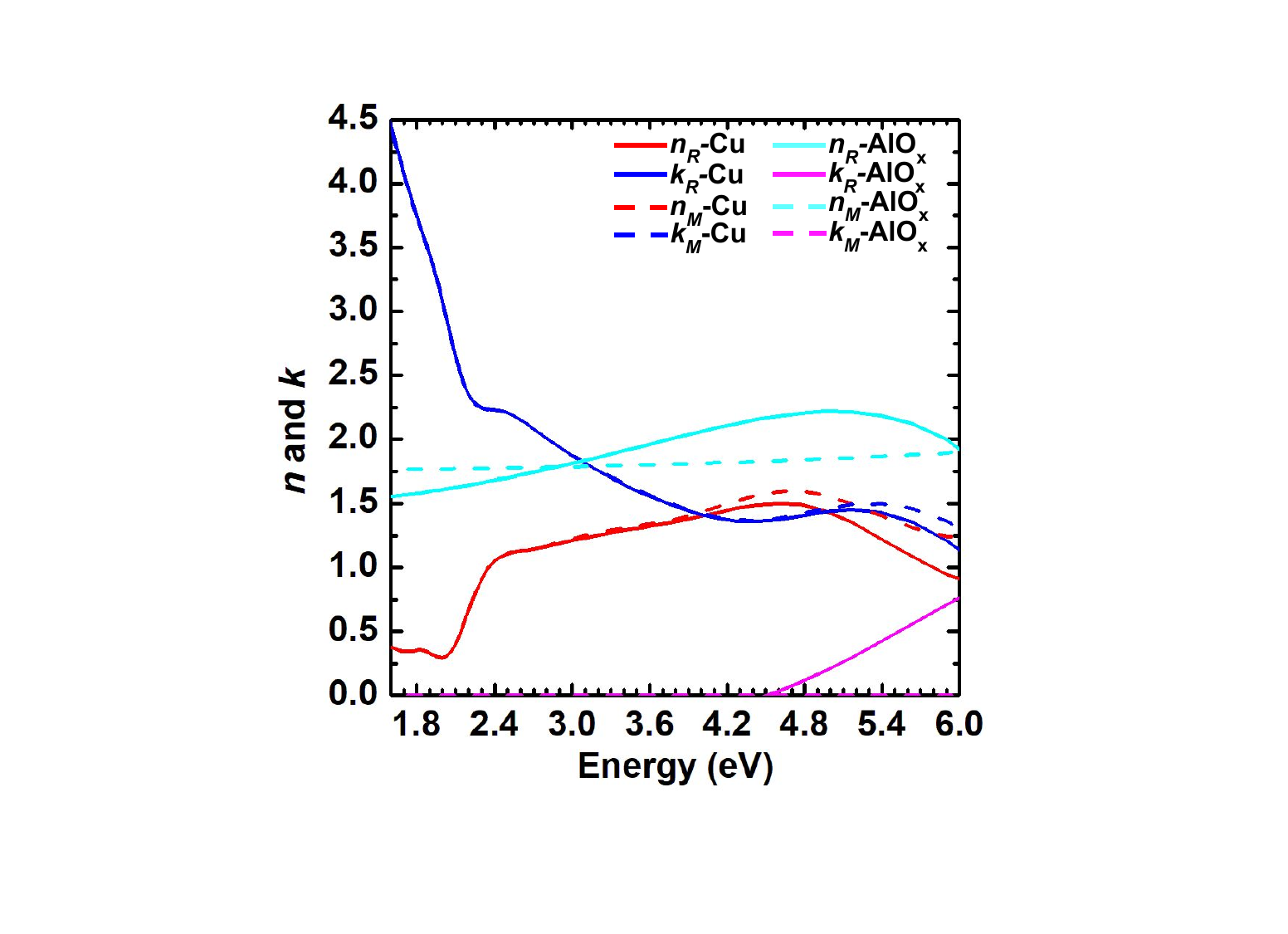}
    \captionsetup{labelformat=empty}\caption{}
    \label{fig:proofConcept}
\end{figure}

 Once the calibration for the two constituent materials – Cu and AlO\textsubscript{x} – was completed as described earlier, dispersions of refractive index $n$ and extinction coefficient $k$ were reconstructed from spectral reflectance of Cu and AlO\textsubscript{x} and plotted in \figref{proofConcept}, providing a proof that the dispersion method worked. In \figref{proofConcept}, solid blue and red curves represent reconstructed $n$ and $k$ dispersions of Cu while dotted blue and red curves depict those of representative Cu found in literatures \cite{QuerryCu}. Solid cyan and magenta curves represent reconstructed n and k dispersions of AlOx while dotted cyan and magenta curves depict those of representative AlO\textsubscript{x} \cite{CRC} \cite{MatrlSci}. The use of representative Cu and AlOx as benchmarks offers a comprehensive characterization of the effectiveness of the dispersion method implemented. Evidently, an excellent match is obtained for Cu in the range of energy pertinent to our study, which is followed by marginal divergence that emerges as energy increases. The divergence is likely to be related to the amount of data available to be used in the calculations, as occurs closer to the ends of the available data range which has no impact on the lower energy range of interest due to an excess of data points. Additionally, the upper bound is heavily influenced by the integration constants – the anticipated error factor – that had to be introduced to perform the integration over the wide energy range.

\noindent \par In contrast to $n$ and $k$ of Cu that show an excellent fit at approximately $99\%$ accuracy over the range of energy of interest in \figref{proofConcept}, the cyan and magenta curves of AlO\textsubscript{x} agree with those of representative AlO\textsubscript{x} found in literatures \cite{SALAD} at approximately $85\%$ accuracy, which is semi-quantitatively consistent with the foundation of the dispersion method that requires the assumption that the material in question exhibits dispersion of complex refractive index that is holomorphic within a domain of energy of interest. Since the imaginary part (i.e., $k$) of the complex refractive index of AlO\textsubscript{x} is essentially zero across the energy range of interest, the imaginary part is not considered holomorphic. As a consequence, the use of the dispersion method for AlO\textsubscript{x} to reconstruct $n$ and $k$ from spectral reflectance leads to $n$ and $k$ that are less accurate than those obtained for Cu. However, our error analysis indicates that errors associated with the integration coefficients of AlO\textsubscript{x} (i.e., errors that arise from the discrepancies seen in $n$ and $k$ of AlO\textsubscript{x} in \figref{proofConcept}) has insignificant contributions to reconstruction of $n$ and $k$ of the short-period AlO\textsubscript{x}-Cu multi-layered nanocomposites samples for which dispersions of complex refractive indices are holomorphic. In addition, the deviation from being holomorphic should be lessened as the percent mass of Cu in the samples, $\tau$, increases and the contribution from Cu begins to dominate. Although the presence of these errors, the rationale of using the dispersion method thus is to obtain distinctive insights of the short-period AlO\textsubscript{x}-Cu multi-layered nanocomposite samples in terms of their material properties by extracting their correlative $n$ and $k$ – optical constants that cannot be measured – from the spectral reflectance shown in \figref{SALAD_graphs}.

\noindent \par
\begin{figure} [h]
    \centering     
    \includegraphics[page=5, trim=0cm 0cm 0cm 2cm, clip, width=1\textwidth]{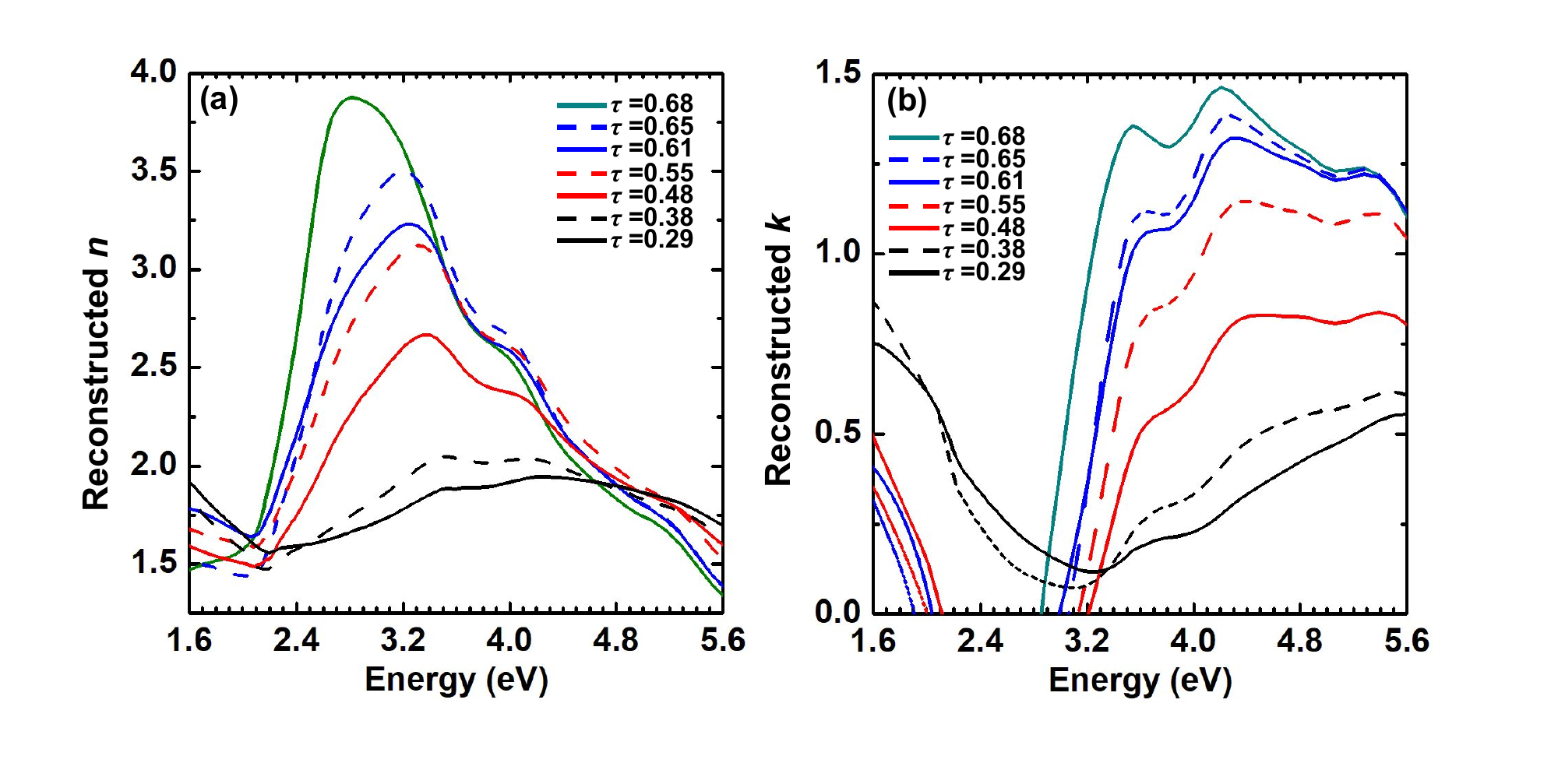}
    \captionsetup{labelformat=empty}\caption{}
    \label{fig:DispResults}
\end{figure}

 Figures 5(a)(b) show the reconstructed dispersions of $n$ and $k$ from the spectral reflectance presented in \figref{SALAD_graphs}. There are a few notable features seen in \figref{DispResults}(a). First, $n$ of the samples with $\tau$ larger than $0.48$ is much larger than those of either Cu or AlO\textsubscript{x} in the range of energy from $2.4$ to $\sim 4.2\,eV$. This is striking in the sense that $n$ of the samples made available by distinctively integrating AlO\textsubscript{x} and Cu does not seem to be expressed by simply interpolating those of the two constituent materials. In addition, each reconstructed $n$ spectra of the samples with $\tau$ larger than $0.48$ in \figref{DispResults}(a) exhibits a peak that shifts depending on $\tau$, while $n$ of AlO\textsubscript{x} and Cu illustrated in \figref{proofConcept} changes monotonically without any peaks. In addition, an additional peak appears to convolute the high energy side of the main peak, which could not have been predicted from the two constituent materials alone. Overall, it is surprising that the $n$ spectra characteristics of the metal-dielectric SALAD samples show key features commonly observed only in the $n$ spectra of semiconductors \cite{OptProp}.

\noindent \par In \figref{DispResults}(b), $k$ appears to fall below zero in the range of energy from $\sim 2.0\,eV$ to $\sim 3.2\,eV$ for $\tau$ greater than $0.38$, which interestingly occurs where their respective absorption due to SPP is expected to emerge as shown in \figref{SALAD_graphs}. The potential explanation for this behavior, previously dubbed regional dominance, predicts that a holomorphic AlO\textsubscript{x} would be dominated by $k$ in a region akin to the derivative of the range of energy dominated by the $n$ region of AlO\textsubscript{x}. For example, in \figref{DispResults}(b), the $k$ values are minimized within the range of energy ($2.0 \sim 2.4,eV$) that matches the energies of the inflection points of the $n$ spectra in \figref{DispResults}(a), which appears to be a property shared only with semiconductors \cite{OptProp}. Moreover, in a similar way to semiconductors, the $n$ spectra reach their peaks at around $3.0 \sim 3.2\,eV$, and then, they reach their localized minimum at around $\sim 3.6\,eV$ at which the $k$ spectra reach their peaks. Further assessment, both electrical and optical, is required to uncover details of underlying physics.

\noindent \par The $k$ spectra in \figref{DispResults}(b) also display a few distinctive features especially for the samples with $\tau$ equal to or greater than $0.48$. All the samples with $\tau$ equal to or greater than $0.48$ exhibit a general feature of the presence of a broad peak that appears to be a convolution of three peaks located at approximately $3.6\,eV$, $4.2\,eV$, and $5.4\,eV$, which casts marked differences when compared with the $k$ spectrum of Cu that decreases monotonically as the energy increases and that of AlO\textsubscript{x} that is essentially zero across the range of energy shown in \figref{proofConcept}. In addition, the $k$ spectra of the samples with $\tau$ equal to or greater than $0.48$ show a drastic decrease when the energy is smaller than $3.0\,eV$, which may indicate the presence of absorption edge associated with optical bandgaps similar to those generally observed in semiconductors.

\noindent \par In general, the optical constants in the inter-band transition region of semiconductors depend on their respective electronic energy-band structures via dielectric functions. For instance, the dielectric functions of a crystalline solid is often approximated by invoking a sum over a finite number of the Lorentz oscillators \cite{dmgGaAs}. These oscillators would individually contribute to and/or collectively manifest as the features seen in \figref{DispResults}(a)(b). Furthermore, the joint-density of states – one of the key factors that determine the shape of $n$ and $k$ spectra – generally show strong variations when the energy approaches those often referred to as critical points \cite{OpPropGaAs}\cite{TempDielect} which is also capable of contributing to unusual, reconstructed $n$ and $k$ spectra presented in \figref{DispResults}.

\noindent \par Another interesting feature present in the SALAD samples is that two local minima (i.e., those at $2.2\,eV$ and $4.2\,eV$ \cite{SALAD} seen on the spectral reflectance of Cu appear to have corresponding local maxima on the spectral reflectance for $\tau\,=\,0.68$ in \figref{SALAD_graphs} – the inversion in spectral reflectance. From this observation comes another explanation to the observed semiconductor-like characteristics; AlO\textsubscript{x} and Cu take turns having their properties dominate with the thickness of the Cu affects the overall magnitude of the reflectance response. For example, it could be said that AlO\textsubscript{x} becomes dominant in the range of energy from $\sim 1.6\,eV$ to $2.3\,eV$, reducing $n$ to the local minima presented in \figref{SALAD_graphs}. Similarly, the immediate increase below $2.0\,eV$ and into the infrared range could be due to Cu rapidly dominating over the AlO\textsubscript{x}. The ultraviolet to visible spectrum may have a more evenly distributed mix of codominance between the materials with Cu beginning to dominate again towards $6\,eV$. It could be this codominance between the constituent materials that leads to the semiconductor-like properties being predicted by the dispersion model. Most importantly we can state that a structure that consists of thin interlaced layers of Cu – a metal – and AlO\textsubscript{x} – a dielectric material – would behave like a semiconductor at least in the visible-ultraviolet spectral range. This of course then implies that through SALAD a new means of producing, at least optically speaking, a new class of semiconductor has been created.

\section{Conclusion}

\noindent \par The purpose of the transfer-matrix method was to simplify extremely complex calculations otherwise needed to determine the spectral reflectance of a multi-layered thin film structure. The transfer-matrix method – a popular method used in modeling optical coatings – failed to produce the spectral reflectance presented in \figref{SALAD_graphs}, which is highly likely to be due to the various assumptions related to the samples that is now expected to be invalid. In contrast, the dispersion method designed to recover complex refractive index from spectral reflectance successfully revealed unique optical features of the samples (i.e., spectral reflectance measurements are all that would be needed to gain significant insight into materials’ optical properties when detailed structural information is hardly obtainable. The implemented model showed considerable promise with the Cu benchmark but struggled slightly with the AlO\textsubscript{x} benchmark. Clearly, there is room for improvement in the dispersion method by, for instance, finding an alternate means to process non-holomorphic materials or make adjustments to the existing dispersion model so that  materials with negligible $k$ in a wide spectral range are appropriately handled. That said, the model does indeed give a glimpse into the nature of the short-period AlO\textsubscript{x}-Cu multi-layered nanocomposites samples deposited by SALAD, revealing that thin interchanging layers of metal and dielectric would behave like semiconductors, at least, in the limited optical domain and providing opportunities in designing and characterizing new optical materials with highly complex structures.

\newpage

\section*{Figure Captions}

\noindent \textbf{Figure 1:} Spectral reflectance of the short-period AlO\textsubscript{x}-Cu multilayered nanocomposite samples, with various $\tau$, prepared by SALAD.
\newline\newline
\noindent \textbf{Figure 2:} Depiction of the short-period AlO\textsubscript{x}-Cu multi-layered nanocomposite sample used in the modeling based on the transfer matrix method. The magnitude of the electric field of incident light (plane-wave) is $E_{in}$. $E_{01}$ symbolizes the magnitude of electric field of an early exiting light wave while $E_{on}$ denotes a later stage exiting light wave that has been reflected to the top from the interface between the first and the second layer. The image illustrates how the incident light is repeatedly transmitted and reflected between layers.
\newline\newline
\noindent \textbf{Figure 3:} The transfer-matrix method was used to calculate the reflectance of spectra of thin-film structures that imitate the short-period AlO\textsubscript{x}-Cu multi-layered nanocomposite samples.
\newline\newline 
\noindent \textbf{Figure 4:} The reconstructed n and k of Cu, $n_R$-Cu (solid red) and $k_R$-Cu (solid blue), and the reconstructed n and k of AlO\textsubscript{x}, $n_R$-AlO\textsubscript{x} (solid cyan) and $k_R$-AlO\textsubscript{x} (solid magenta), are plotted as a function of energy. $n_R$-Cu, $k_R$-Cu, $n_R$-AlO\textsubscript{x}, and $k_R$-AlO\textsubscript{x} are compared with their respective measured n and k, $n_M$-Cu (dashed red), $k_M$-Cu (dashed blue), $n_M$-AlO\textsubscript{x} (dashed cyan), and $k_M$-AlO\textsubscript{x} (dashed magenta), found in literature. $n_M$-Cu and $k_M$-Cu are from Querry \cite{QuerryCu} and $n_M$-AlO\textsubscript{x} and $k_M$-AlO\textsubscript{x} are from the Sellmeier equation \cite{CRC}\cite{MatrlSci}.
\newline\newline
\noindent \textbf{Figure 5:} The reconstructed n in \textbf{(a)} and k in \textbf{(b)} of the AlO\textsubscript{x}-Cu nanocomposite thin-film samples with varying percent copper m.

\newpage
\fontsize{30pt}{12pt}{\selectfont \noindent FIG 1}
\vspace{4cm}

\newpage
\fontsize{30pt}{12pt}{\selectfont \noindent FIG 2}
\vspace{4cm}

\newpage
\fontsize{30pt}{12pt}{\selectfont \noindent FIG 3}
\vspace{4cm}

\newpage
\fontsize{30pt}{12pt}{\selectfont \noindent FIG 4}
\vspace{4cm}

\newpage
\fontsize{30pt}{12pt}{\selectfont \noindent FIG 5}
\vspace{6cm}

\newpage
\printbibliography

\end{document}